\documentclass[aps, prd, twocolumn, superscriptaddress, nofootinbib, showpacs]{revtex4}
\usepackage[dvips]{graphicx}

\def\lsim{\lower0.6ex\vbox{\hbox{$ \buildrel{\textstyle <}\over{\sim}\ $}}}
\def\gsim{\lower0.6ex\vbox{\hbox{$ \buildrel{\textstyle >}\over{\sim}\ $}}}

\def\beq{\begin{equation}}
\def\eeq{\end{equation}}

\def\bfig{\begin{figure}[t] \begin{center}}
\def\efig{\end{center} \end{figure}}

\def\Om{\Omega_{\rm M}}
\def\Ol{\Omega_{\rm \Lambda}}

\def\rhocrit{\rho_{\rm crit}}

\def\himpc{{h$^{-1}$ Mpc}~}

\def\etal{{\it et al.}~}
\def\cf{{\it cf.}~}
\def\eg{{\it e.g.},~}
\def\ie{{\it i.e.},~}

\def\mnras{{Mon. Not. R. Astron. Soc.}~}

\def\4he{$^4$He}

\def\7li{$^7$Li}
\def\8Be{$^8$Be}

\def\Msun{$M_{\odot}$~}

\def\V{V(\phi)}
\def\Mvir{M_{\rm vir}}
\def\Rvir{R_{\rm vir}}
\def\cvir{c_{\rm vir}}
\def\Vvir{V_{\rm vir}}
\def\Vmax{V_{\rm max}}
\def\dvir{\Delta_{\rm vir}}
\def\rs{r_{\rm s}}
\def\dv2{\Delta_{\rm V/2}}
\def\Rv2{r_{\rm V/2}}
\def\Mcoll{M_*(z_c)}
\def\rhoc{\tilde\rho_{\rm s}}
\def\sig8{\sigma_8}

\begin{document}

\title{Inflation, cold dark matter, and the central density problem}

\author{Andrew R. Zentner}
\email{zentner@pacific.mps.ohio-state.edu}
\affiliation{Department of Physics, The Ohio State University,
Columbus, OH 43210, USA}
\author{James S. Bullock}
\email{james@astronomy.ohio-state.edu}
\affiliation{Department of Physics, The Ohio State University, 
Columbus, OH 43210, USA}

\date{\today}

\begin{abstract}

The lingering problem  with high central densities  in  dark halos has
arisen in the context of (L)CDM cosmologies with $n=1$ scale-invariant
initial power spectra.  Although $n=1$ is often justified by appealing
to the inflation  scenario,  the choice  is not  generally  justified.
Specifically, inflation models with mild but important deviations from
scale invariance ($n \sim  0.9$)  are  not uncommon,  and  those  with
significant ``running'' of the   spectral index are quite   plausible.
Even a mild deviation  from scale invariance  can be important because
halo collapse  times and densities  depend  on the relative amount  of
small-scale   power.   Here,   we  choose    several   popular,  often
well-motivated, models of inflation and work out the ramifications for
galaxy  central densities.      For  each model,   we  calculate   its
COBE-normalized primordial power spectrum  and deduce the implied halo
densities   using a semi-analytic  method   calibrated  against N-body
simulations.  We compare our predictions to a sample of $\sim 50$ dark
matter-dominated   galaxies  using  a  non-parametric   measure of the
density, $\dv2$,  defined  as the mean mass  density,  relative to the
critical density, within the radius at  which the rotation curve falls
to half of  its maximum value.   While  standard $n=1$ LCDM halos  are
overdense  by   a factor  of    $\sim 6$,    several of   our  example
inflation+CDM  models predict   halo densities well  within,  and even
below, the  range preferred  by  observations.  We  also show  how the
presence of  massive  ($m_{\nu} \sim 0.5$  eV)  neutrinos can help  to
alleviate the central  density  problem, even with a   scale invariant
spectrum.   We conclude that  galaxy central densities  may  not be as
problematic for the CDM paradigm  as is sometimes assumed: rather than
telling us something about the nature  of dark matter, galaxy rotation
curves may be telling  us something about inflation  and/or neutrinos.
An important  test of this  idea will be  an eventual consensus on the
value of $\sigma_8$, the  rms overdensity on  the scale $8 h^{-1}$Mpc.
Our successful models tend to have  values of $\sigma_8 \approx 0.75$,
which is well  within  the range  of recent determinations.   Finally,
models with $n > 1$ (or $\sigma_8 \gsim 1$) are highly disfavored.
\end{abstract}


\maketitle

\section{\label{sec:intro}Introduction}

The standard model  of structure formation (LCDM) is  one in which the
universe is  dominated by cold, collisionless dark  matter (CDM), made
flat  by  a cosmological  constant  ($\Lambda$),  and endowed with
initial   density  perturbations   via  quantum   fluctuations  during
inflation  \cite{BFPR}.   Although   the  need  for  the  cosmological
constant  component  is  unexpected,  the  CDM+inflation  paradigm  is
strongly  motivated, and with  the parameter  choices $\Om  = 1  - \Ol
\approx  0.3  -  0.5$,  $h  \approx  0.7$, LCDM  can  account  for  an
impressive range   of   astronomical  observations on  large   scales.
However, on galactic scales, this model faces some potentially ruinous
difficulties.  Perhaps the most  troubling of  these problems are  the
indications that  the central regions of  galaxies are noticeably less
dense than the favored LCDM model typically predicts \cite{first,cden,
B95,  dslope,  MQGSL99,ABW01,BMR01,BB02,keeton}.    In this paper   we
suggest  that these    observations  might  be telling us something
fundamental about the epoch of inflation, and also explore how
massive neutrinos affect halo densities.

There are  two related   but  distinct facets of the   central density
problem.  The first problem concerns the fact that the integrated mass
densities within well-defined  central radii of observed galaxies (see Sec. 
\ref{sub:model}) seem to  be a  factor  of $\sim  6-8$  larger than 
the central  densities predicted  by  standard LCDM \cite{cden,ABW01,keeton}.  
The second, often referred  to as the ``cuspy  halo problem,'' highlights the fact
that CDM  halo density profiles  are  predicted to diverge  at small r 
($\rho  \propto  r^{-\alpha}, \alpha   \sim  0.7-1.5$),  while  galaxy
rotation curves are often better fit with constant density cores
\cite{first,dslope,B95,MQGSL99,BMR01,BB02}.     While   the second of
these  issues has received the most  attention, it is often degenerate
with    the first problem and    is  somewhat more controversial.  For
example, de Blok and collaborators \cite{BB02}  have argued that their
sample  of  low-surface  brightness  galaxies  favor  fits  to density
profiles  with a constant  density central  ``core''  over those  with
cusps; however, van  den Bosch and Swaters   argue that a  majority of
galactic  rotation curves are  acceptably fit by divergent profiles as
long  as they are much less centrally concentrated than typical
halos  in    standard LCDM   \cite{vdBS}.    Furthermore,   {\em  all}
observational errors ({\it e.g.}, slit offset)  tend to favor constant
apparent central density over cusps.  At present, it is not clear that
the cuspy halo problem presents  a serious challenge to LCDM, although
it appears that   the data do prefer halos   that are less   centrally
concentrated than typical halos in the standard LCDM model.

The problems with central densities  have triggered a growing  concern
that we  are  missing something fundamental   in our understanding  of
galaxy  formation.  (This is in   spite of the  fact  that some of the
problematic claims are disputed   \cite{vdBS,KZS,Jimenez,lowenstein,rines}.)
Solutions to the problems range from those that use baryonic physics
\cite{KZS,navarro,katz,dekel,Bullock01,wechsler}
to those  that rely   on  altering the   nature   of the  dark  matter
\cite{SIDM,SFDM,wdm,LHZB,BKW}.  While the astrophysical solutions are
reasonably  well-motivated, the fact that problems seem
to exist for relatively small, 
dark matter-dominated dwarf galaxies all the way to large ellipticals suggests
that a single baryonic solution may not be able to address all of our concerns.
The altered dark matter solutions, on  the other hand, could  possibly
be made to  match  the range of  observations, but only by invoking
unmotivated or fine-tuned candidates (there is no well-motivated 
warm dark matter candidate).  

Our inflation-derived solution is nonastrophysical but  works entirely 
within the desirable tenets of the CDM paradigm.  This paper is conservative 
in that it concentrates on the integrated density within some well-defined 
radius, which is certainly more robustly determined in simulated halos than 
the central slope of the density profile.  As we shall discuss below, 
the same is likely true for observed galaxies.  

Our  work  is  principally  inspired  by Alam,  Bullock, and Weinberg
\cite{ABW01} who  suggested that the central density  problem would be
reduced  significantly  in  LCDM  if the  initial  inflationary  power
spectrum were  tilted to favor  large scales.  The term  ``tilted'' is
defined  in  terms  of   the  primordial  power  spectrum  of  density
fluctuations, which we assume over some range in wave number $k$ can be
written as  $P(k) \propto k^n$,  corresponding to a mass  variance per
logarithmic interval of $\Delta^2(k)=k^3P(k)/2\pi^2$.  Tilted  power
spectra refer  to those with $n  \ne 1$.  In ``standard''  LCDM, it is
assumed that  $n$ is exactly $1$, corresponding  to a scale invariant,
Harrison-Zel'dovich primordial  power spectrum.  This  choice is often
justified by the tendency for inflation models to predict {\em nearly}
scale invariant spectra;  however, generic models of  inflation do not
predict   primordial   power    spectra   that   are   {\em   exactly}
scale invariant.  As  we discuss below, the central  densities of dark
halos are extremely  sensitive to the amount of  small-scale power and
hence small deviations from scale invariance can be very important.

Our  aim is  to compute  the predicted  primordial $P(k)$  for several
example inflation  models and apply  these results to the  question of
galaxy central densities and concentrations.  Although similar in spirit 
to the agenda of Kamionkowski and Liddle \cite{KL00}, who suggested that 
the small-scale crisis  might  reflect  a  sharp  feature in  the  
inflationary  power spectrum (see Sec.  \ref{sub:BSI}), our mind-set 
is to  look at models that  are  not  particularly  fine-tuned.   We  
simply  choose  fairly
representative,  simple,  single  field  inflationary  models  and  we
examine  a range  of predicted  power spectra,  even one  with  with a
``blue'' initial spectrum  ($n>1$).  In the context of slow roll inflation, 
models that predict significant
tilt generally yield   effective spectral  indices that are   strongly
scale-dependent  or  exhibit significant  ``running''  of the spectral
index.   Consequently, when comparing observational  data  that span a
wide range of scales ($\Delta \ln k \gsim 12$  in this case) it makes
sense to account for the variation of $n(k)$ with scale in addition to
the so-called ``tilt'' of the spectrum.  We account for the running of
the spectral index by calculating it in specific inflationary models and 
show that the running can have an important effect on structure on 
galactic scales.  We also estimate the  effect of a
``hot dark matter'' component in  the form of massive neutrinos on the
central densities  of dark  matter halos.  We  make our  estimates for
halo densities  using a semi-analytic model  normalized against N-body
simulations.  The model relies on our, now well-founded, understanding
that  halo  central densities  are  determined by  their collapse  
histories \cite{wechsler}.

Although we examine models  with varying amounts of small-scale power,
we are  not free  to alter the spectrum by an  arbitrary amount.
The amplitude of power on small ($ \sim 8$ \himpc) scales  
is often quantified in terms of $\sig8$, which is the rms overdensity smoothed with
a top-hat filter of radius $8$ \himpc.  Observationally, this quantity
can be determined  in a number of ways  including techniques that rely
on the  abundance of rich x-ray  clusters, the cosmic  shear from weak
gravitational lensing, and galactic peculiar velocity flows.  However,
these estimates  do not converge on  a definitive value (even  when the
same method  is used by  different authors) and many  recent estimates
seem  to  advocate surprisingly  low  values  of $\sig8$  \cite{MSVNL,
SIG8}.  Roughly speaking, recent estimates yield values that span the
range $0.55 \lsim \sig8 \lsim 1.2$  for $0.3 \lsim \Om \lsim 0.5$.  In
the  following, we  will  only  consider models  with  $\sig8 >  0.55$
because models  with a smaller  $\sig8$ do not  have a good  chance of
being able to match observations.  We acknowledge that even this limit
is pushing the  observational bounds  (although it  is consistent
with Ref.  \cite{MSVNL}) but we  feel that it  is best to  explore all
possibilities for  the sake of alleviating the  tension between theory
and   observation   on   subgalactic   scales.    For   reference,   a
scale invariant spectrum ({\it i.e.},  with spectral index $n=1$) that
is normalized  to the  Cosmic Background Explorer (COBE) measurements
\cite{COBE}  of  the  large-scale  cosmic microwave  background  (CMB)
anisotropy  via   the  fitting  forms  of  Bunn,   Liddle,  and  White
\cite{BLW}, has $\sig8 \simeq 0.95$ assuming that the gravitational 
wave contribution is negliglible.

The remainder of this paper is organized as follows.  In Sec. 
\ref{sec:pot} we introduce several models of inflation and
calculate the power spectra predicted by each model.  We give a 
short description of the effects of massive neutrinos on 
the evolved linear power spectrum in 
Sec. \ref{sec:neutrinos}.  We discuss the properties of 
dark matter halos and describe our semi-analytic model 
for estimating halo central densities in Sec. \ref{sec:CEN}.  
In Sec. \ref{sec:results} we present our results and compare
them with the observed central densities of dwarf and 
low surface brightness (LSB) galaxies.  Lastly, we summarize 
our conclusions and indicate directions for future work in 
Sec. \ref{sec:con}.  

Throughout this work we will assume
$\Omega_{\rm M} = 0.3$, $\Omega_{\rm \Lambda} = 0.7$, 
$\Omega_{\rm B}h^2 = 0.020$, and $h = 0.72$.

\section{\label{sec:pot}Inflationary Power Spectra}

It  is widely believed that  the primordial density perturbations that
led  to the growth of  structure in the  Universe were produced during
inflation: quantum fluctuations  in the inflaton  field were frozen in
as the rapid  cosmological  expansion stretched these  fluctuations to
length  scales   larger  than the   horizon.    The  power spectrum of
primordial  perturbations   can  be  calculated   via   the  slow roll
approximation    (for a review,  see  Ref.  \cite{LR99} and references
therein).   The standard calculation,  to  lowest order in  slow roll,
yields expressions for the spectrum of density perturbations at 
horizon crossing, 

\beq
\label{eq:deltaH}
\delta_{\rm H}^2(k) \simeq 
\frac{1}{75\pi^2 m_{pl}^6}\frac{V^3(\phi)}{V'^2(\phi)},
\eeq
and the effective spectral index of the primordial spectrum,
\beq
\label{eq:spectind}
n(k) \simeq 1 + 2\eta - 6\epsilon
\eeq
in  terms of the inflaton potential  $\V$ ($V'(\phi) \equiv dV/d\phi$,
$V'' \equiv d^2V/d\phi^2$)   and  the slow roll parameters   $\epsilon
\equiv m_{pl}^2/2(V'/V)^2$  and  $\eta \equiv  m_{pl}^2(V''/V)$.   The
reduced  Planck  mass is  defined in   terms  of Newton's constant  as
$m_{pl} \equiv 1/\sqrt{8\pi G_{\rm  N}} \simeq 2.4 \times 10^{18}$ GeV
and, as usual,  these  expressions  are to   be evaluated  at  horizon
crossing (\ie when $k=aH$).

In the limit of exact de Sitter space during inflation, the predicted 
primordial power spectrum would approach exact scale invariance; however, 
any model in which a scalar field is slowly rolling towards a minimum of its 
potential will predict some deviation from $n=1$.  Of course, 
in   the  context  of some  inflationary  models, the deviation is
quite small, and scale invariance is a reasonable approximation.  
One frequently cited example of this sort is power-law inflation, for
which there is an  exact solution  \cite{AW84}.  The reason why approximate 
scale invariance is expected in this model has to do with estimates of
the gravitational wave contribution to the CMB  quadrupole.  In addition to
scalar  density fluctuations, inflation  also produces tensor (gravity
wave) fluctuations.  In the power-law inflation case, the ratio of the
tensor  to  scalar  contribution to the  CMB   quadrupole, 
$r \equiv C_2^{\rm tensor}/C_2^{\rm scalar}$,
increases with  the tilt as $r \simeq  6.9(1-n)$. 
A similar result also applies 
to chaotic   inflation   models    (\eg Ref. \cite{Linde83}) because 
$\phi \sim m_{pl}$ in these models.  Recent CMB measurements indicate that 
the tensor contribution is small ($r \lsim 0.2$  \cite{WTZ01,Hann01}) 
so power-law inflation requires $n \gsim 0.97$. 

However,   this  case   does   not    exemplify  general  inflationary
predictions.  Tensor perturbations can  be negligible even if the tilt
is not.  This is because the tensor wave amplitude depends only on the
energy scale of inflation.  The gravity wave contribution
is negligible if the inflaton field remains far below the Planck 
scale, as would be expected in well-motivated models such as 
the running-mass case discussed below.  Models meeting this requirement 
can  naturally produce modest tilts and spectral index running.  
Moreover, there are reasonably well-motivated cases that can yield dramatic 
departures from scale invariance.

In the balance of this section, we  briefly outline the predictions of
several models of inflation that lead to  deviations from the standard
$n=1$,  scale invariant  primordial spectrum   and  present the  $z=0$
linear power spectra in each case.  Included in  this set of models is
a  more     extreme  example   that     exhibits  so-called   ``broken
scale invariance'' and for which the slow roll approximation cannot be
used.  In all other cases, we  calculate the primordial power spectrum
to  second order  in slow roll using  the  method of  Stewart and Gong
\cite{SG01}   which   is  sufficiently    accurate   for  our purposes
\cite{GL96}.  In this way, we  take into account both  the tilt of the
power spectrum and the  running of the spectral  index.  

To derive the
low-redshift power  spectra, we use  the fitting form for the transfer
function given by Eisenstein and Hu \cite{EH99} and the exact relation
for the linear growth  factor in flat  cosmologies with a cosmological
constant given by Bildhauer,  Buchert, and Kasai \cite{BBK92}.  In all
cases we  normalize   the power spectrum  to  COBE  using  the fitting
functions of Bunn, Liddle  and White \cite{BLW}.  We consider  several
models in which the effective spectral index varies significantly with
scale.  In these cases, we follow  the prescription of Ref. \cite{BLW}
and evaluate the normalization at the scale $k_* = 7H_0 \simeq 0.0023$
h Mpc$^{-1}$, which is approximately the pivot scale of the COBE data,
using the effective spectral index at that scale, $n(k_*)$.

\subsection{\label{sub:IPL}Inverted power law potentials}

We begin with the illustrative example of the inverted 
power law (IPL) potential (or ``small field polynomial'' in the 
language of Ref. \cite{Hann01}) which has the basic 
characteristics of ``new inflation'' \cite{ST84}.  The general form is 

\beq
\label{eq:IPL}
\V = V_0(1-c\phi^p)
\eeq
with $p>2$.  This potential implies that 
the effective spectral index of the primordial power spectrum 
on the scale $k_*$ is given by 

\beq
\label{eq:n_IPL}
n(k_*) \simeq 1-2\Bigg(\frac{p-1}{p-2}\Bigg)\frac{1}{N_*}, 
\eeq
where $N_*$ is the number of e-folds of inflation that occur 
between the epoch when the scale $k_*$ leaves the horizon and the 
end of inflation.  We can obtain a rough estimate of $N_*$ in 
terms of the energy density at reheating, $\rho_{\rm RH}$, the value of 
the inflaton potential when $k_*$ leaves the horizon, $V_*$, and the value 
of the inflaton potential at the end of inflation, $V_{\rm F}$, by 
assuming instantaneous transitions between vacuum domination and matter 
domination at the end of inflation and matter domination and radiation 
domination at reheating.  This gives 

\beq
\label{eq:Nstar}
N_* \approx 57 - \ln\Bigg(\frac{10^{15} \textrm{ GeV}}{V^{1/4}_*}\Bigg)
+  \ln\Bigg(\frac{V^{1/4}_*  \rho_{\rm  RH}^{3/4}}{V_{\rm F}}\Bigg)  .
\eeq  
If the  details  of the  end  of inflation  and  the process  of
reheating were  known, $N_*$ would be known  precisely; however, these
details are  not known.  In  order to obtain definite  predictions, we
take $N_* = 50$ which  is a fairly standard working hypothesis.  Using
this in Eq. (\ref{eq:n_IPL}), we see that with $p=4$ (we refer to this
model   as   IPL4)   this   model   predicts   mild   deviation   from
scale invariance,  namely  $n(k_*)  \simeq  0.94$.   Accordingly,  the
spectral  index  is  mildly scale-dependent,  $|dn(k)/d\ln(k)|  \simeq
0.001$.  Figure 1 depicts a typical  power spectrum at  $z=0$ with the
choice  $p=4$.  Rather  than $P(k)$  or $\Delta^2(k)$,  we plot  the rms
overdensity on  a given  mass scale $\sigma(M)$,  because this  is the
relevant  quantity   for  our  subsequent   calculations.   The  COBE
normalization amounts to choosing  a suitable combination of $V_0$ and
$c$ and  the effective spectral  index is insensitive to  this choice.
Normalized to  COBE, this model predicts a  perfectly acceptable value
of   $\sig8  \simeq  0.83$.\footnote{The   length  scale   $8$  \himpc
corresponds to a mass scale  of $M \simeq 1.8 \times 10^{14}$ h$^{-1}$
\Msun under the assumption that $\Omega_{\rm M} = 0.3$.}

Before proceeding, we mention that the particle physics motivation for
this type  of potential  may be somewhat  dubious.  In  particular, if
$p=4$, COBE normalization requires the dimensionless $\phi^4$ coupling
constant to  be of  order $10^{-14}$ (the  fine-tuning problem  may be
obviated by considering the coupling to be a parameter of an effective
field  theory  rather   than  a  fundamental  parameter)  \cite{LR99}.
Nevertheless,  we consider  this  to be  a  good illustrative  example
because it is simple and  has the general behavior $|n(k)-1| \simeq {\cal
O}(1)/N(k)$ that  is exhibited by  a wide variety of  models including
many specific incarnations of new inflation \cite{ST84},  hybrid inflation 
and  mutated hybrid inflation \cite{Stewart95},  as well  as  models 
with  a variable  gravitational constant  \cite{SBB89}.   In   addition, 
this  potential  mimics  the potential encountered  in a  particular 
variation of mutated hybrid inflation known as ``smooth'' hybrid 
inflation \cite{LPV96}.

\subsection{\label{sub:RM}Running-mass model}

Stewart has proposed a model in  which the need to fine-tune the inflaton 
mass in order for inflation to occur in the context of supergravity is 
eliminated  by a ``flattening'' of the  effective inflaton potential due  to  
loop corrections  \cite{Stewart97}.
This  provides a  natural mechanism  for generating  a  potential that
gives  rise  to  inflation.   Interestingly, the  resulting  effective
potential  can lead to  a spectral  index considerably  different from
$n=1$ and  with a significant scale  dependence.  In this  model it is
assumed that  in the  sector of the  inflaton field,  supersymmetry is
broken  explicitly during inflation  and the  scalar fields  have soft
supersymmetry-breaking  mass  terms  as would generally be the case.
Through couplings  to fields with  soft supersymmetry-breaking masses,
the scalar  field masses may get  important renormalization corrections.
The  one-loop  correction to  the  inflaton  potential  then gives  an
effective potential with a running inflaton mass,

\beq
\label{eq:RMpot}
\V \simeq V_0 + \frac{1}{2}m^2(\phi)\phi^2 + \dots, 
\eeq
where the ellipsis represents nonrenormalizable terms that become 
important at the Planck scale.   The value of $V_0$ is tied
to the scale of the supersymmetry breaking during inflation,  
$M_{\rm S} \sim V_0^{1/4}$.

This type of model has been discussed extensively by Covi, Lyth and Roszkowski \cite{CLR99}
and Covi and Lyth \cite{CL99} who derived cosmic microwave background 
constraints on such models.  Inflation occurs in the vicinity of 
an extremum of the potential and the established blueprint for 
analyzing inflation in the context of this model is to assume 
that $m^2(\phi)$ can be approximated by a function that is linear 
in $\ln(\phi)$ while cosmological scales are leaving the horizon \cite{Stewart97, 
CLR99, CL99}.  In the notation of Refs. \cite{CLR99, CL99}, the effective 
potential is then written as in Eq. (\ref{eq:RMpot}) with $m^2(\phi) \simeq 
-(V_0/m_{pl}^2)(c/2 - c\ln(\phi/\phi_0))$ and the spectral index in this case 
is given by

\beq
\label{eq:RM_n}
n(k_*) \simeq 1 + 2\sigma e^{-cN_*} - 2c \eeq where $\phi_0$ is chosen
such that  $V'(\phi_0) = 0$ and  $c$ and $\sigma$  are parameters that
may be either  positive or negative and we  generally expect that $|c|
\lsim |\sigma| \lsim 1$ \cite{CLR99,CL99}.   With $c>0$ and $\sigma < 0$, we
have the  particularly interesting case that $n<1$  and decreases with
increasing $k$.   With $c<0$ and $\sigma>0$, $n>1$  and decreases with
scale.   We use  such  a model  in  several places  to illustrate  the
predictions  of a  model with  a  primordial power  spectrum that  has
$n>1$.  For  reasonable parameter choices  there may be  a significant
tilt and  the scale  dependence may be  as strong  as $|dn(k)/d\ln(k)|
\simeq 0.005$  on cosmologically interesting  scales.  The significant
running of the  tilt is not surprising.  As we  stated earlier, in the
context  of   slow roll  inflation,  models   with  significant  tilts
typically  exhibit strong variation  of $n$  with scale.
The COBE  normalization is related  to the parameters $V_0$  and $\tau
\equiv  |c|\ln(m_{pl}/\phi_0)$.   In  each  case  we  make  physically
reasonable   choices  of   these  parameters   to  enforce   the  COBE
normalization   (see  Ref.  \cite{CL99}).  The shape of the spectrum is 
relatively insensitive  to these  choices. We choose  this model  as a 
simple  example of  an  inflationary model  that  may {\em  naturally}
predict significant deviations from  $n=1$ and running of the spectral
index.

In Fig. 1 we show present-day, linear power spectra for two particular
choices of parameters.  The model with $n>1$ (RM $n>1$) has parameter
choices $\sigma = 0.05$ and $c = -0.001$, resulting in $n(k_*) \simeq
1.1$ and $\sig8 \simeq 1.21$, and is shown largely for illustration
(note that in this case a hybrid mechanism is necessary to end
inflation).  The more interesting case with $n<1$ (RM $n<1$) has
parameter choices $\sigma = -0.31$ and $c = 0.04$.  For this model,
$n(k_*) \simeq 0.84$ and $|dn/d\ln(k)| \simeq 0.004$ which is
consistent with constraints on tilt from various analyses of CMB,
large-scale structure and Ly$\alpha$ forest data \cite{WTZ01,Hann01}.
For this model $\sig8 \simeq 0.65$ which is on the low side
of our acceptable range.

\subsection{\label{sub:BSI}Spectra with broken scale invariance}

In contrast with the above models, there may be a feature at some scale 
in the inflationary potential that causes the power to drop abruptly.  This 
possibility leads us to consider models with so-called 
``broken scale invariant'' (BSI) spectra.  In these models, there is a
critical scale $k_c$, and  for  $k \gg k_c$  and $k  \ll  k_c$ the   primordial power
spectrum has an effective  power law index $n  \simeq 1$.  However, on
scales near  the critical scale the amplitude  of the  initial density
perturbations  changes abruptly so that the  power on scales $k > k_c$
can be significantly less than that on scales $k < k_c$.  This type of
spectrum may  arise in models in  which  more than  one field plays an
important role in inflation  while cosmological scales are leaving the
horizon   \cite{multifield} but   placing  the    scale $k_c$   in  an
observationally  interesting  range  usually introduces  a fine-tuning
issue.

As an idealized case of BSI, Starobinsky derived an analytic 
expression for the primordial power spectrum in a model where the inflaton 
potential has a step discontinuity in its first derivative 
(\cf Eq. [\ref{eq:deltaH})] \cite{Starobinsky92}.  This is a useful model to study 
for two reasons.  First, the primordial power spectrum can be written in a 
relatively simple closed form.  Second, and more material, 
this model exhibits the most rapid drop in power possible in a single field model of 
inflation \cite{KL00}.  Lesgourgues, Polarski and Starobinsky \cite{LPS98} investigated 
using primordial power spectra of this type to explain a feature on scales of 
about $125$ h$^{-1}$ Mpc in the galaxy power spectrum measured by the Automatic 
Plate-measuring Machine survey \cite{GB98} (such a feature is not present in more recent 
determinations of the galaxy power spectrum) while Kamionkowski and Liddle \cite{KL00} 
explored the effects of such a primordial spectrum on the typical abundance of 
dwarf satellites.   

\begin{figure}
\resizebox{!}{9cm}
{\includegraphics{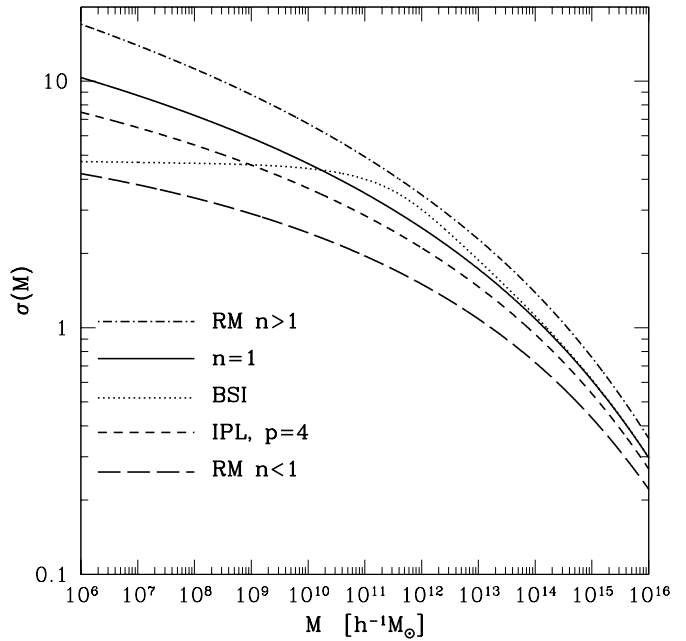}}
\caption{\label{fig:spectra}
The $z=0$ rms overdensity as a function of 
mass scale predicted by several models of inflation and normalized to 
COBE.  The models depicted here are the running mass model with 
$\sigma = 0.05$ and $c = -0.001$ (RM $n>1$), the $n=1$ scale invariant 
Harrison-Zel'dovich spectrum ($n=1$), the broken-scale invariant model 
with $k_{c} = 0.9$ and $p=10$ (BSI), the inverted power law model with 
index $p=4$ (IPL4), and the running mass model with $\sigma = -0.31$ 
and $c = 0.04$ (RM $n<1$).}
\end{figure}

In this scenario, the power spectrum of density perturbations $\Delta_p^2(k)$, 
prior to being modified by causal physical processes, is given by the following exact 
relation \cite{Starobinsky92}: 

\begin{eqnarray}
\label{eq:BSI_pps}
\Delta_p^2(k) \propto y^4\Bigg[1-3(p-1)\frac{1}{y}\Bigg(f_{-}\sin(2y) 
+ \frac{2}{y}\cos(2y)\Bigg) \nonumber \\
+ \frac{9}{2}(p-1)^2\frac{1}{y^2}f_{+}\Bigg(f_{+} + f_{-}\cos(2y) - \frac{2}{y}\sin(2y)\Bigg)\Bigg]
\end{eqnarray}
where $y\equiv  k/k_c$, $f_{\pm} \equiv 1  \pm 1/y^2$, and $p$  is the
ratio  of the amplitude of   fluctuations on scales $k   < k_c$ to the
amplitude of fluctuations on scales $k >  k_c$.  In this model, we are
free to choose the amplitude of primordial tensor perturbations and we
assume that they  are negligible.   The normalization is according to
the  COBE data.  Inspired  by  the  work  of  Kamionkowski and Liddle
\cite{KL00}, we choose $p=10$ and $k_c =  0.9$ h Mpc$^{-1}$  in order
to suppress power on mass scales $M  \lsim 10^{10}$ h$^{-1}$ \Msun and
thereby alleviate the  dwarf satellite  problem  (note that  we have
chosen a different  $k_c$ than Kamionkowski   and Liddle \cite{KL00},
partly  because   we assume a  different  cosmological  model).   The
dotted line  in Fig. \ref{fig:spectra} shows  the  $z=0$, linear power
spectrum  predicted by this model.  As there is a rise in power prior
to    the   cutoff   at $M    \simeq    10^{10}$ h$^{-1}$ \Msun  (see
Fig. \ref{fig:spectra}),  we find that 
  $\sig8 \simeq 0.97$ which is
slightly larger than the value in a scale invariant model. 

\section{\label{sec:neutrinos}Massive Neutrinos}

A  preponderance  of  evidence  from solar  and  atmospheric  neutrino
oscillation experiments  like Super-Kamiokande \cite{SK},  the Sudbury
Neutrino   Observatory   \cite{SNO},   the  Russian-American   Gallium
Experiment \cite{SAGE},  the Gallium Neutrino  Observatory \cite{GNO},
the  Gallium  Experiment  \cite{GALLEX},  and  the  Soudan  Experiment
\cite{SOUDAN}  seems to  imply  that neutrinos  are indeed massive.  Yet these 
experiments cannot determine the absolute magnitude of the neutrino masses 
and it may be that the masses are large enough to have significant 
cosmological implications.  If massive neutrinos
(or  other ``hot dark matter''  particles) make  up  a non-negligible
portion of the dark matter, the effect of their free-streaming will be
to reduce power relative to the standard model on small length  scales.  
This situation is commonly referred to as the cold+hot dark matter 
scenario \cite{primack}.

It is easy  to estimate the
scale at which this  effect becomes important.  Massive neutrinos will
move at a speed over order $c$ until they  become nonrelativistic when
$m_{\nu}  \sim 3T_{\nu}$  which occurs  at a  redshift of  $z_{\rm NR}
\simeq 2 \times 10^3 (m_{\nu}/\textrm{eV})$  and we expect power to be
suppressed  on  scales smaller  than  the  horizon  scale at  redshift
$z_{\rm NR}$.  As such, a rough  estimate is that power will be damped
on all scales $k \gsim k_{\rm FS}$ where
\beq
\label{eq:k_fs}
k_{\rm FS}                           
\simeq
0.03\Omega_{\rm M}^{1/2}\sqrt{\frac{m_{\nu}}{\textrm{eV}}}  
\textrm{h Mpc}^{-1}.
\eeq
This  corresponds to  suppression of  power  on mass  scales $M  \lsim
M_{\rm      FS}      \simeq      3      \times      10^{18}\Omega_{\rm
M}^{-3/2}(m_{\nu}/\textrm{eV})^{-3/2}$     h$^{-1}$     \Msun.     The
contribution  of  $N_{\nu}$  massive,  light  ($m_{\nu}  \ll  1$  MeV)
neutrinos  to  the  mean  matter  density, relative  to  the  critical
density,              is              $\Omega_{\nu}             \simeq
N_{\nu}(m_{\nu}/\textrm{eV})h^{-2}/92$.   On  scales $k>>k_{\rm  FS}$,
the fractional suppression of power due to massive
neutrinos approaches a value of $\sim (1+8\Omega_{\nu}/\Omega_{\rm M})^{-1}$ 
\cite{MBDGS}.

This  modification to  the power  spectrum  on small  scales has  been
studied in  detail by many authors (\eg Refs.  \cite{MBDGS, EH99}).
In  fact,  the  current  best  bounds on  neutrino  masses  come  from
demanding consistency of  the power spectrum on scales  probed by COBE
and the  smaller scales probed  by clusters \cite{FLS00}  and the
Ly$\alpha$  forest \cite{CHD99}  or  from the  shape  of the  observed
galaxy power spectrum on scales $0.01 \lsim k \lsim 0.2$ \cite{2DF02}.
Roughly  speaking,  these cosmological  bounds  dictate that three nearly 
mass-degenerate neutrinos must have $m_{\nu}\lsim 1-2$ eV \cite{primack}
while direct bounds on the  electron neutrino mass from tritium decay 
experiments give $m_{\nu_e} \lsim 3$ eV \cite{TRIT}.

\begin{figure}
\resizebox{!}{9cm}
{\includegraphics{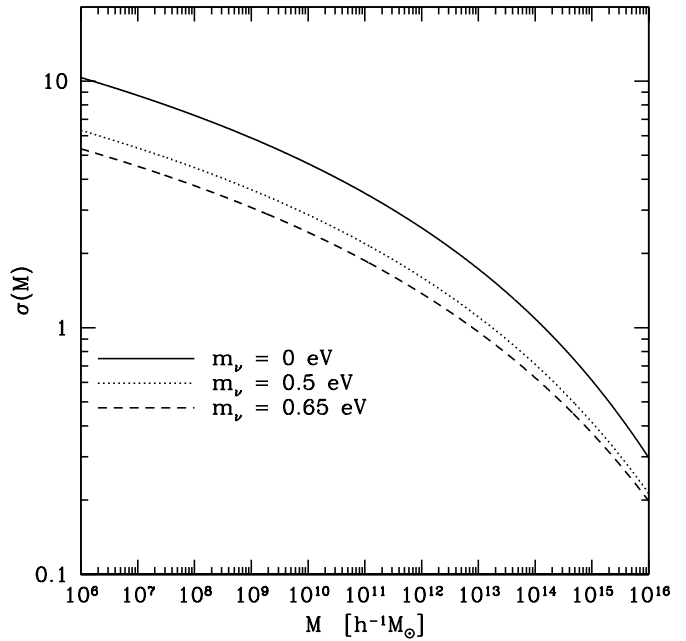}}
\caption{\label{fig:spectnu} Power spectra with massive neutrinos compared 
to the standard, scale invariant power spectrum with no massive neutrinos.}
\end{figure}

In  what  follows,  we  study   the  effect  of  the suppression  of
small-scale power by three neutrinos with effectively degenerate 
masses $m_{\nu} \lsim 1$ eV on the central  densities  of dark  matter  
halos.   As  was pointed  out  in Ref.   \cite{FLS00},  excessively  
large   neutrino  masses   lead  to unacceptably low  values of  $\sig8$.  
Our strategy  is to fix  $\Om = 1-\Ol = 0.3$ and $n=1$ and to ascertain 
whether or not a neutrino mass
that saturates  our lower  limit of $\sig8  > 0.55$ can  alleviate the
central densities problem associated  with the CDM paradigm.  For this
cosmology, the lower limit on  $\sig8$ is saturated by a neutrino with
$m_{\nu} =  0.65$ eV.   For comparison, we  also report results  for a
model with $m_{\nu} = 0.5$ eV  which has $\sig8 \simeq 0.64$.  We show
in Fig. \ref{fig:spectnu}, the present-day linear power spectra of the
scale invariant  reference  model  and  the two  models  with  massive
neutrinos.   Notice the  suppression  of power  on  all relevant  mass
scales.

\section{\label{sec:CEN}The Central Densities of Dark Matter Halos }

\subsection{\label{sub:desc}Describing dark matter halos}

The absolute size of a virialized dark matter halo can be described by
the virial mass $\Mvir$, or equivalently the virial radius $\Rvir$, or
the virial velocity $\Vvir^2  \equiv G\Mvir/\Rvir$.  The virial radius
of a  halo is defined as the  radius within which the  mean density is
equal to the virial overdensity $\dvir$, times the mean matter density
of the Universe, $\rho_{\rm M}$.  Thus $\Mvir$ and $\Rvir$ are related
by

\beq
\label{eq:mvir}
\Mvir = \frac{4\pi}{3}R_{\rm vir}^3\rho_{\rm M}\Delta_{\rm vir}.
\eeq
The  virial  overdensity is  set  by  the  spherical top-hat  collapse
approximation  and  for  flat  cosmologies  the value  of  $\dvir$  at
redshift $z$ can be approximated by \cite{BN98}

\beq
\label{eq:dvir}
\dvir(z) \simeq \frac{18\pi^2 + 82x - 39x^2}{x+1},
\eeq
where $x+1 \equiv \Omega_{\rm M}(z) = \Omega_{\rm M}(1+z)^3/(\Omega_{\rm M}(1+z)^3 + 
\Omega_{\rm \Lambda})$.  With $\Omega_{\rm M} = 0.3$ and $\Omega_{\rm \Lambda} = 0.7$, 
$\dvir \simeq 337$ at $z=0$.  

Several  analytic density  profiles for  dark matter  halos  have been
proposed  as good  approximations  to the  results of  high-resolution
N-body  simulations   and  all  agree   at  large  radii.    Moore  et
al. \cite{MQGSL99} found that the density in the central region of the
halo varies  as $\rho  \propto r^{-1.5}$ and  so proposed  the density
profile

\beq
\label{eq:moore}
\rho_{\rm M}(r) = \frac{\rho_{\rm s}}{(r/r_{\rm M})^{1.5}(1+r/r_{\rm M})^{1.5}}
\eeq
while Navarro, Frenk, and White \cite{NFW} (hereafter NFW) found that density in the central 
regions of halos varies as $\rho \propto r^{-1}$ and have therefore proposed the profile 

\beq
\label{eq:NFW}
\rho_{\rm NFW}(r) = \frac{\rho_{\rm s}}{(r/r_{\rm s})(1 + r/r_{\rm s})^2}.
\eeq
In  most of what follows, we  will assume an  NFW profile.  For the NFW
profile the  two  parameters  are  related   by $\rho_{\rm  s}  \simeq
\rho_{\rm  NFW}(0.466r_{\rm s})$.  If   N-body simulations do  predict
profiles somewhere between the $\rho \propto r^{-1}$ and $\rho \propto
r^{-1.5}$, the behaviors of the NFW and  Moore profiles respectively, then
the  NFW profile  is a  conservative choice  in  the sense that  Moore
profiles predict more centrally concentrated halos \cite{ABW01}.  
NFW profiles can more easily match the data and require 
less drastic modifications of the standard paradigm in order to do so.

A useful criterion for assessing the relative central concentration of
a halo is the concentration parameter

\beq
\label{eq:defcvir}
\cvir \equiv \frac{\Rvir}{\rs}. 
\eeq
Although this  quantity is defined explicitly in  terms of a parameter
of the NFW   profile, viz. $\rs$, there  is   no significant loss   of
generality because  we  can identify  $\rs$ with   the  point at which
$d\ln\rho/d\ln r = -2$ and thereby relate these results to the results
obtained using another density profile \cite{Bullock01}.  
Restating the NFW profile as a velocity curve gives 

\beq
\label{eq:NFWV}
V^2_c(r) = \Vvir^2\frac{\cvir}{f(\cvir)}\frac{f(x)}{x}.
\eeq
with $x \equiv r/\rs$ and $f(y) \equiv \ln(1+y) - y/(1+y)$.  The maximum 
velocity is given by 

\beq
\label{eq:NFWVmax}
V_{\rm max}^2 \simeq 0.216 \Vvir^2 \frac{\cvir}{f(\cvir)}
\eeq
and occurs at  a radius $r_{\rm max} \simeq  2.16\rs$.  

A more directly observable measure of halo central densities has 
been proposed by Alam, Bullock, and Weinberg \cite{ABW01}.  This 
quantity is known as the ``central density parameter'' and is 
defined as 

\beq
\label{eq:dv2def}
\dv2 \equiv \frac{\overline \rho (\Rv2)}{\rho_{\rm crit}} = 
\frac{1}{2}\Bigg(\frac{\Vmax}{H_0 \Rv2}\Bigg)^2
\eeq
or the mean dark  matter density  within the  radius $\Rv2$ where  the
galaxy rotation  curve goes to half of  $\Vmax$.   In practical units,
$\dv2  \simeq    5 \times    10^5  (\Vmax/100 \textrm{    kms}^{-1})^2
(\Rv2/\textrm{h}^{-1}\textrm{   kpc})^{-2}$.  Assuming an NFW profile,
$\Rv2 \simeq 0.13\rs$ and the central density parameter is given by 

\beq
\label{eq:dv2nfw}
\dv2^{\rm NFW} \simeq 340\cvir^3/f(\cvir). 
\eeq

There are distinct advantages to using $\dv2$ to characterize the 
central densities of dark halos.  For one, $\dv2$ is more robustly 
determined, both observationally and in numerical data from N-body 
simulations, than is the inner slope of a density or velocity 
profile yet it probes scales small enough to betray the conflict 
between theory and observation.  Moreover, $\dv2$ is defined without 
reference to any particular density or velocity profile.

\subsection{\label{sub:model}Predicting halo central densities}

To find  the central densities of  dark matter halos  predicted by the
aforementioned inflationary  models, we make use of the semi-analytic
model of  Bullock \etal \cite{Bullock01} who were stimulated by the previous
work of NFW \cite{NFW}.  This model has been calibrated to
the results of high-resolution N-body  simulations and although the model was 
developed in the context of scale invariant  CDM power  spectra, \footnote{
The Bullock et al. model was shown to work well in predicting the redshift
and mass dependence of halo profiles for a standard LCDM model, and
also reproduces the $z=0$ results presented by NFW for standard CDM,
open CDM, LCDM, and  several power-law models \cite{Bullock01}.}
it was shown  to work well in predicting the results  of an LCDM simulation with 
$n=0.9$, as discussed  in Ref. \cite{ABW01}.  This model represents 
an improvement over the previous NFW model because it reproduces the 
relationship between $\cvir$ and $\Mvir$ observed in N-body simulations as 
a function of redshift whereas the NFW model fails at $z>0$.  
It is important 
to  realize however, that  the following treatment is simplified  
and untested over
the full range of power spectra  we apply it to.  Specifically,
this model has not been tested against simulations with running spectral
indices nor has it been tested against simulations with a significant hot 
dark matter component.  Note, however, that halos formed within hot + cold dark 
matter simulations do seem to follow an NFW profile down to $\sim 2\%$ of 
the halo virial radius \cite{KKBP}.  Ideally, our results
will motivate future  work using N-body simulations in  order to test
our preliminary conclusions.  

Briefly, the model of Bullock \etal \cite{Bullock01} (and NFW) embodies the  fact  
that we  expect the  central densities  of dark matter  halos to  reflect 
the  mean density  of the Universe at a  time when the central region of  
the halo was accreting matter at a high rate \cite{wechsler}.  Therefore, 
we expect halos with central regions that collapsed earlier to be denser than 
their late-forming counterparts.  Accordingly, our first step is to assign an 
epoch of collapse to a halo via the prescription that, at the collapse epoch $z_c$, 
the typical collapsing mass, $M_*(z_c)$, is equal to some fixed fraction 
of the halo's virial mass.  Explicitly, we define  

\beq
\label{eq:coll}
\Mcoll \equiv F\Mvir.
\eeq
$\Mcoll$ is the mass scale at which the rms density fluctuation is equal 
to the equivalent linear overdensity at collapse, $\delta_c \simeq 1.69$.  
If  $\sigma(M,z)$ is the rms  overdensity on
mass scale $M$  at redshift $z$  [we use $\sigma(M)$  with no redshift
argument to denote $\sigma(M,z=0)$ as usual],  then the collapse criterion  
can be written  as  $\sigma(M_*,z_c) = 1.69$.  Notice that this definition of 
the epoch of collapse differs from that of NFW who defined the collapse epoch 
using the extended Press-Schechter formalism \cite{EPS}.  This is the key 
difference that gives the improved model the ability to trace the redshift 
dependence of the $\Mvir-\cvir$ relationship.  $F$ is a free parameter and Bullock 
\etal found that the model is in good  agreement with the results of N-body 
simulations if $F  = 0.01$ \cite{Bullock01}.  The small value of the
parameter $F$ is not surprising.   The densities that characterize the
very  central regions of  halos are determined  by the power on scales
much smaller than  the size of the  halo, scales that broke  away from
the expansion at a much earlier time than the mass scale $\Mvir$.

It is already evident that  the central densities of dark matter halos
will be  very sensitive to $\sigma(M)$  on small scales  and hence, to
the slope of the primordial power spectrum or the presence of hot dark
matter.  At early times (but during matter domination), $\Omega_M \sim
1$, and $\sigma(M,z) \propto  (1+z)^{-1}$.  Thus the epoch of collapse
varies  approximately  as $(1+z_c)  \propto  \sigma(F\Mvir)$.  If  the
central  densities  do, in  fact,  reflect  the  mean density  of  the
Universe at  the epoch of  collapse then, roughly speaking,  we expect
$\dv2 \propto \sigma^3(F\Mvir)$ so that  a change in power by a factor
of $2$ will lead  to a change in central density by  a factor $\sim 8$
[\cf Eq. (\ref{eq:dv2nfw}) and Eq. (\ref{eq:cvir}) below].

The second step is to relate the mass density of the Universe at $z_c$
to a  characteristic halo density  today.  Bullock \etal chose  to use
$\rhoc$ defined by

\beq
\label{eq:rhodef}
\Mvir \equiv \frac{4\pi}{3}\rs^3\rhoc.
\eeq
For an NFW profile, $\rhoc = 3\rho_{\rm s}f(\cvir)$.  Introducing the free 
parameter $K$, we associate $\rhoc$ with the universal density at collapse 
via

\beq
\label{eq:Kden}
\rhoc = K^3\rhocrit\dvir(z=0)\Omega_{\rm M}(1+z_c)^3.
\eeq
Solving Eqs. (\ref{eq:rhodef}) and (\ref{eq:Kden}) for $\cvir$ gives 

\beq
\label{eq:cvir}
\cvir(\Mvir) = K(1+z_c(\Mvir)).
\eeq
Agreement  with N-body  simulations fixes $K   = 4.0$.  Bullock \etal
and Wechsler \etal claim that the $1\sigma$ scatter in  
the $\cvir-\Mvir$ relation is roughly $\Delta \log(\cvir) \simeq 0.14$ 
while Jing has   argued  for a somewhat   smaller scatter  given  by $\Delta
\log(\cvir) \simeq 0.08$ at $1\sigma$ \cite{JING}.  

With this model in place, we can use the linear power spectra of 
the previous two sections to predict $\cvir$ and, more practically,
$\dv2$ and compare them with observation.

\section{\label{sec:results}Results}
  
In this section we compare the predictions of the previous models with
data on  the rotation   curves of   dwarf  galaxies  and  low  surface
brightness (LSB) galaxies.  We  concentrate our discussion on galaxies
with both HI and H$\alpha$ data or HI data that has been corrected for
the  effects of beam  smearing.  The  data we  use are taken  from the
recent  works of Swaters \cite{S99}  (mass modeling  of these galaxies
was  performed   by van den Bosch  and   Swaters \cite{vdBS}) de Blok,
McGaugh and  Rubin \cite{BMR01} and de  Blok and Bosma \cite{BB02} who
combined previous HI measurements \cite{S99, PHI} with high-resolution
H$\alpha$  measurements.  We use  these  data to  derive observational
estimates of $\dv2$ for  comparison with the theoretical predictions.  For
the data of de  Blok and Bosma  \cite{BB02}, we use their best fitting
model for the dark  matter distribution of each  galaxy in the absence
of  baryons to derive estimates for  $\dv2$.   For the data of Swaters
\cite{S99} and de Blok, McGaugh,  and  Rubin \cite{BMR01}, we fit  the
raw  data to  the   velocity   profile  proposed by  Kravtsov    \etal
\cite{KKBP},
\beq
\label{eq:kkbp}
V_{\rm c}(r) =
V_c^0\frac{(r/r_k)^{\gamma}}{[1+(r/r_k)^{\alpha}]^{(\gamma+
\beta)/\alpha}}, 
\eeq
and  use the  best fitting  models  to estimate
$\dv2$.   The  profile  in   Eq.  (\ref{eq:kkbp})  has  the  practical
advantage  that  it  parameterizes  the sharpness  of  the  transition
between the two power laws at large and small radii.  Hence the fitted 
value  of the  effective power  law index  at small  radii is  to some
degree decoupled  from the details of  the rotation curve  at $r \gsim
r_k$.   This added  versatility makes  it a  very useful  and accurate
formula for  describing observed rotation curves at  small radii.  Our
estimates of $\dv2$ are robust in  that for most galaxies in the above
samples, the inferred values of $\dv2$  change by less than 40\% if we
instead fit the data  with NFW, pseudoisothermal or Burkert \cite{B95}
density profiles.\footnote{It is interesting that
there is no systematic difference in the derived $\dv2$
from one profile to the next.  However,
Moore profiles tend  to fit the data 
more poorly, and give larger variation in the implied $\dv2$.
This is similar to the result found 
by  van  den  Bosch  and  Swaters  \cite{vdBS}.}   The
robustness of  the central density parameter is  yet another advantage
of  using $\dv2$  as a  diagnostic of  the central  densities  of dark
matter halos.

Any     comparison of the      predictions  of  N-body  simulations  or
semi-analytic calculations that  model  the behavior  of CDM with data
rests  on  some assumptions  about the  physics of  baryon infall.  In
making this comparison, we  believe that our methods  are conservative
in the   sense that   we  likely  overestimate  $\dv2$ based  on   the
observational data  in  order to give the   data every  opportunity to
match theoretical predictions (including the scale invariant
``standard  model'').  First, we restrict ourselves   to dwarf and LSB
galaxies  which are generally believed to  be dominated  by their dark
matter  components \cite{S99, BM97}.   In so doing,  we  hope that any
effects of baryonic infall are mitigated  but recognize the fact that we
may be introducing a heretofore unappreciated selection effect.  Second,  
we calculate $\dv2$ based on the full rotation curve without mass modeling or 
baryon subtraction.  We therefore overestimate  the central density of 
the primordial dark matter halo because the cooling and contraction of 
the baryons likely lead to contraction of the dark matter component as well  
(see Ref. \cite{first}, although also see Ref. \cite{katz}).  
Third, the measured rotation curves of about 20\% of the galaxies 
in the sample may not extend out to large enough radii for an
accurate determination of $\Vmax$ and consequently $\Vmax$ may be
significantly underestimated for several galaxies.  In these cases, we
simply take the  last point in the rotation curve as an  estimate of
$\Vmax$.  By examining Eq. (\ref{eq:dv2def}) it is easy to see that if
$V_{\rm c}(r) \propto r^{\gamma}$ with $\gamma \le  1$ at small radii,
an {\em underestimation} of  $\Vmax$ by a factor  $f_{\rm v}$ leads to
an {\em  overestimation} of  $\dv2$  by a factor of $f_{\rm
v}^{2(1-1/\gamma)}$  (clearly, for $\gamma   = 1$, corresponding to  a
constant density core, the  error cancels exactly).  In other  words,
the error   introduced  has the  net   effect of  bringing  theory and
observation closer together.

\begin{figure}
\resizebox{!}{8cm}
{\includegraphics{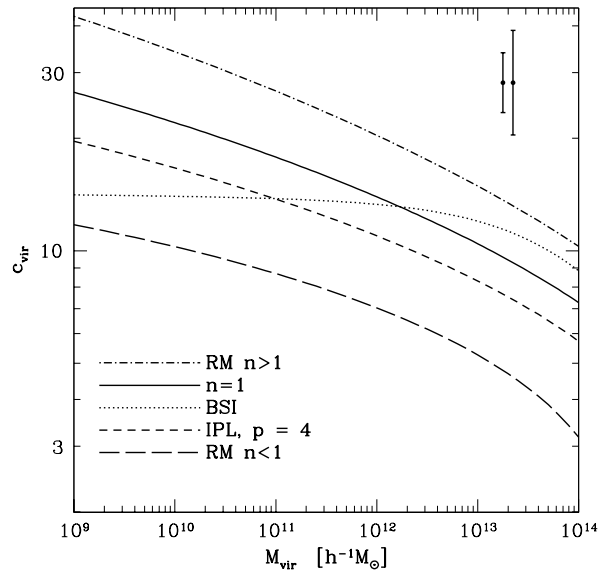}}
\caption{\label{fig:cvirs}The median $\cvir-\Mvir$ relation predicted by several 
different primordial  power spectra.  The predictions corresponding to
the different primordial power spectra are labeled in the same fashion
as in Fig. \ref{fig:spectra}.  Bullock \etal \cite{Bullock01} estimate
the $1\sigma$   scatter to be $\Delta   \log(\cvir) \simeq 0.14$ while
Jing argues for a smaller  scatter of $\Delta \log(\cvir) \simeq 0.08$
\cite{JING}.  These estimates for the $1\sigma$ scatter are illustrated 
by the error bars in the upper right corner.}
\end{figure}

\begin{figure}
\resizebox{!}{8cm}
{\includegraphics{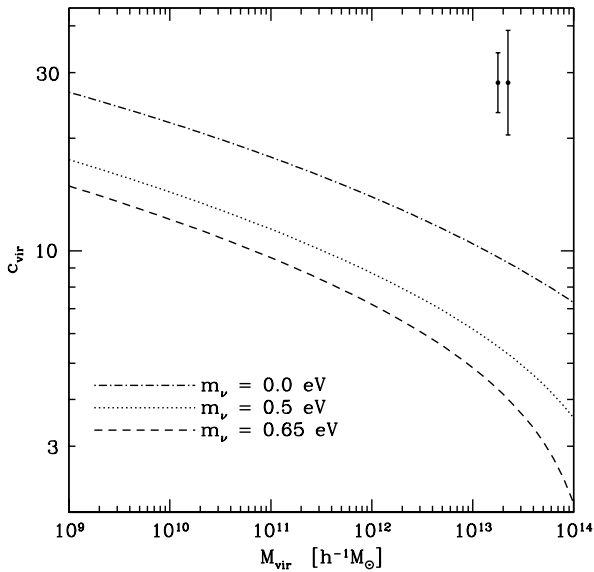}}
\caption{\label{fig:cvirs_nu} 
The median $\cvir-\Mvir$ relation in models 
with massive neutrinos.}
\end{figure}

In   Fig.  \ref{fig:cvirs}, we show   the theoretical predictions  for the
concentration  parameter $\cvir$ in the context of our inflationary models.
Figure \ref{fig:cvirs_nu} shows the    predictions  in scenarios  with
massive  neutrinos.  Notice the wide swath  of the $\cvir-\Mvir$ plane
that  is  carved out  by the various  models and,  in particular, that
$\cvir$  can be  reduced  by  a factor of two or more  by  adopting
primordial power spectra  predicted by reasonable models  of inflation
or by adding neutrino masses that are not ruled out by observation
or experiment.  Dark matter halos may be significantly less concentrated 
than standard LCDM plus scale invariance predicts.

Unfortunately, the $\cvir-\Mvir$ relation  is not  directly observable
and, what is more, it is defined in terms of a particular density profile.  
To connect theory with observations, we compare the quantity $\dv2$, as 
a measure of inner halo concentration, to $\Vmax$ as a measure of the 
absolute size of the halo.  For an NFW profile, $\Vmax$ is related to  
$\Mvir$ through Eqs. (\ref{eq:mvir}) and (\ref{eq:NFWVmax}).

\begin{figure}
\resizebox{!}{9cm}
{\includegraphics{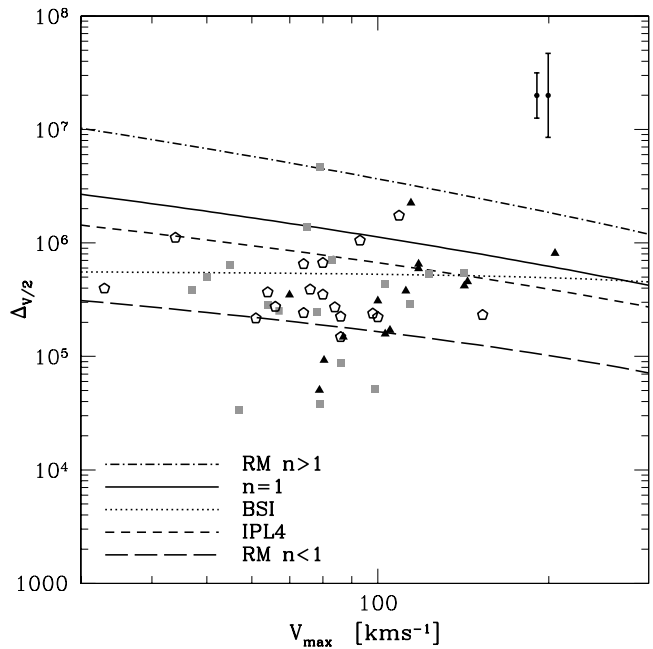}}
\caption{\label{fig:dvcomp}$\Vmax$ vs. $\dv2$ predictions compared with data.  
The  filled triangles are the  data points from  de Blok, McGaugh, and
Rubin \cite{BMR01}, the  gray squares are derived from  the data of de
Blok and Bosma \cite{BB02}, and the open  pentagons are points derived
from the  data  of  Swaters \cite{S99}.   The  different  theoretical
predictions are labeled in the same  way as Fig. \ref{fig:cvirs}.  The
error bars  in the  upper  right corner  show   the expected  $1\sigma$
scatter in the theoretical predictions.  The smaller range corresponds
to the Jing \cite{JING} estimate  and the larger range corresponds  to
the estimate of Bullock \etal \cite{Bullock01}.}
\end{figure}

\begin{figure}
\resizebox{!}{9cm}
{\includegraphics{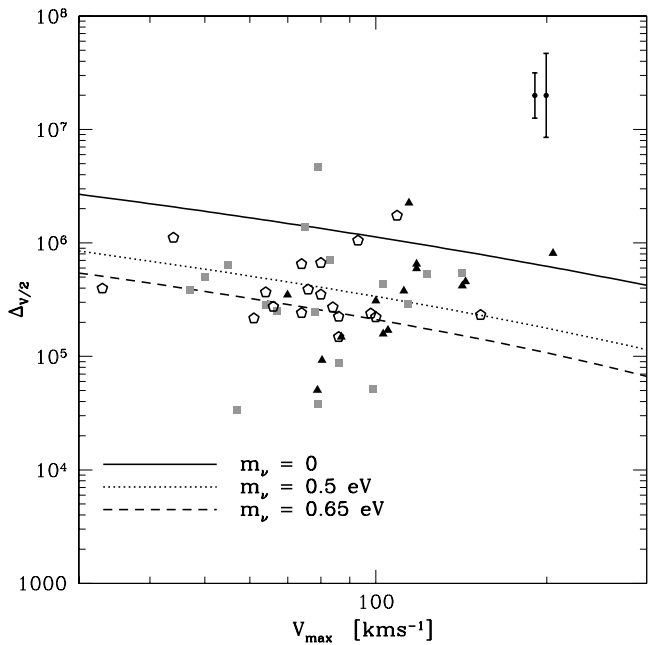}}
\caption{\label{fig:dvcomp_nu} $\Vmax$ vs. $\dv2$ predictions in models with 
massive neutrinos compared with data.  The data points and the error bars in 
the upper right corner are the same as in Fig. \ref{fig:dvcomp}.}\end{figure}

The results of this comparison are shown in Figs. \ref{fig:dvcomp} and
\ref{fig:dvcomp_nu}.  First,  consider the predictions  of the various
models of  inflation.  Although the agreement or disagreement of a 
particular model with the data is hard to quantify, it is not surprising that 
the running mass  model with $n>1$ is effectively ruled out by the data.  
More interestingly,  we find that in agreement    with   previous    studies   
\cite{ABW01}, the $n=1$ scale invariant spectrum also  has difficulty reproducing 
the observed galactic  central densities.  This  is a  restatement of  the problem:
{\em if  we are not  preferentially selecting low density  galaxies by
restricting attention  to low  surface brightness and  dwarf galaxies,
then some additional physics is needed to reconcile the standard model
of  CDM plus  scale invariant  primordial spectrum  with the  observed
central  densities of  dark matter-dominated  galaxies}.  IPL4  does a
somewhat better  job of  matching the data  but the moderate  tilt and
spectral  index running  in this  model are  likely not  sufficient to
bring theory and observation together.  For BSI, the agreement is much
better but note that it is difficult to lower the theoretical $\dv2$ values 
further by adjusting the parameters of the model.  Increasing $p$, the 
ratio of power on scales $k<<k_c$ to power on scales $k>>k_c$, does not 
do much to help the BSI model  come  closer  to  matching  the data  
because  the  fluctuation amplitude cannot drop quickly enough to produce 
a significant decrease in $\sigma(M)$  on scales of  interest.  Meanwhile, 
we  cannot decrease $k_c$  much further without  threatening the  success of  
the standard model on large scales.

Notice  that  the  running-mass  model  with $n<1$  (RM  $n<1$)  is  a
relatively good match to the median of the data in the $\Vmax-\dv2$ plane 
(perhaps even  undershooting  the  median).  It  is  worth  noting  that  this
agreement has  come without the need  to saturate our  lower bounds on
spectral tilt from  CMB and large scale structure  ($n \approx 0.9 \pm
0.1$, see Refs. \cite{Hann01,WTZ01}) or our lower limit on $\sig8$.  The central 
densities of dark matter halos are very sensitive to the initial power spectrum 
and it seems as though  the predicted central densities of  dark matter halos
in a  LCDM cosmology may be  reduced to acceptable  levels by invoking
simple and  well-motivated models of inflation  with $n <  1$ and/or a
running spectral index.

Likewise, in the case of  massive neutrinos, we see that by saturating
our lower bound  on $\sig8$, we can reduce  the predicted median value
of  $\dv2$ to  observationally  acceptable levels.   It  seems that  three
massive  neutrinos with  $ 0.5  \textrm{ eV}  \lsim m_{\nu}  \lsim 0.65
\textrm{  eV}$ can  decrease  small-scale power  enough  to provide  a
pretty good match to the values of $\dv2$ inferred from rotation curve
data.

\section{\label{sec:con}Conclusions and Discussion}

The  central density  problem is  one of  several 
difficulties confronting  the 
standard  paradigm of  structure formation.   In this
paper  we  explored solutions  that do  not  invoke
uncertain baryonic physics but preserve the cold and collisionless
properties  of the dark  matter. In  Sec. \ref{sec:results}  we showed
that models  of inflation  that predict moderate,  yet observationally
acceptable tilts, $ 0.8 \lsim  n \lsim 0.9$, may provide an acceptable
solution to the central density problem.  These tilts are consistent
with  the latest constraints  from joint  analyses of  CMB anisotropy,
large-scale      structure     and     Ly$\alpha$      forest     data
\cite{Hann01,WTZ01,croft}.  Moreover,  these tilts can  be produced in
well-motivated models of inflation.  In fact, we worked in the context
of specific models throughout this paper and in so doing, we were able
to  take into  account  the important  effect  of the  running of  the
spectral index.  To illustrate the  importance of the running, we also
considered a ``tilted''  model with no spectral running  and $n \simeq
0.84$ (the effective tilt of the RM $n<1$ model on the COBE scale) and
found that  this model predicts  central densities that are  more than
$40 \%$ larger than the those predicted by the RM $n<1$ model over the
range $30$ kms$^{-1} \le \Vmax \le 200$ kms$^{-1}$.

Given that precise measurements of the tilt of the power spectrum and 
the running of the spectral index using the data from the Sloan Digital
Sky Survey \cite{SDSST} (URL http://www.sdss.org/) and the Microwave 
Anisotropy Probe (URL http://map.gsfc.nasa.gov/index.html)
are on the horizon \cite{EHT99}, it may be useful to adopt an empirical 
stance and consider the maximum tilt and running that are acceptable with 
respect to galactic central densities without linking the tilt and running 
through a particular inflationary model.  As it is difficult to quantify 
the agreement or disagreement of a particular parameter choice with the 
data and because the current data certainly do not constrain the slope of the 
relationship between $\Vmax$ and $\dv2$, we adopt the somewhat arbitrary, 
but sensible, criterion that a model predicts unacceptably diffuse galaxies 
if $\dv2 \le 10^5$ at $\Vmax = 100$ kms$^{-1}$.  Using the this criterion, 
we    find that a lower limit on  $n(k_*)$ 
allowed as a function of $dn(k_*)/d\ln k$
can be approximated as
\beq
\label{eq:mtilt}
n(k_*) + 6.76dn(k_*)/d\ln k \ \gsim \ 0.77.
\eeq
These maximally tilted models have $\sig8 > 0.55$ for $n(k_*) \gsim 0.75$.  
If we adopt the criterion that a ``good'' fit to the data has $\dv2 
\simeq 3 \times 10^5$ at $\Vmax = 100$ kms$^{-1}$ then a good fit to 
the data is given approximately by 
\beq
\label{eq:gtilt}
n(k_*) + 6.97dn(k_*)/d\ln k \simeq 0.87.
\eeq

We also showed that massive neutrinos with $0.5$ eV $\lsim m_{\nu} 
\lsim 0.65$ eV may provide an alternative solution to the central 
density problem; however, we consider this solution rather less 
attractive.  In order for neutrinos to solve the central density 
problem, it is necessary to nearly saturate our adopted lower 
limit on $\sig8$ because, relative to the standard scale invariant 
model,  the power spectrum is damped by a factor 
$\sim (1+8\Omega_{\nu}/\Om)^{-1}$ on scales smaller than $\sim 10^{16}$ \Msun 
(corresponding, roughly speaking, to $k \gg k_{\rm FS}$) whereas in the 
inflationary models, power falls off continuously with increasing wave number.  
The range of neutrino masses allowed by the above criterion that the dark 
matter halos not be too diffuse is $m_{\nu} \lsim 0.9$ eV, but as we 
mentioned earlier a neutrino mass greater than $\sim 0.65$ eV leads to 
unacceptably small values of $\sig8$.  A neutrino mass of $m_{\nu} = 0.9$ eV 
implies that $\sig8 \simeq 0.46$.

We  did not  deal  directly with  the  issue of  central slopes.   The
problematic  issue here  is  that  cold (and  warm)  dark matter  halo
densities diverge at small radii whereas galactic rotation curves seem
to     be     fit    better     with     constant    density     cores
\cite{first,dslope,B95,MQGSL99,BMR01,BB02}.  While this is a worrisome
situation,  as we discussed above it is  difficult to  tell the  degree 
to  which this  is a serious challenge  to LCDM.  First, {\em all} 
observational errors favor constant density cores over cusps.  Second, while 
it has been  observed that pseudoisothermal density  profiles with constant  
density cores often fit  galactic rotation  curves  better than  NFW profiles 
\cite{cden,BMR01, BB02}, the conclusion  that observations 
indicate halos have cores rather than cusps is a  nonsequitur.  This is 
because {\em all} points on the  curve contribute  to the  fit.  Rotation 
curve fits are often  largely determined by the transition region  between 
the two power laws of the profile,  and may not be faithful representations 
of the observed rotation curves at small radii (where there are relatively few
data points).  In addition, van den Bosch and Swaters showed that most
rotation curves can be acceptably fit by divergent density profiles as
long  as  the  galaxies  are  much less  centrally  concentrated  than
standard LCDM predicts \cite{vdBS}.  To address the inner slope issue,
it  is  probably  more  useful  to  use  a  fitting  form  similar  to
Eq. (\ref{eq:kkbp}),  despite the  fact that it  is not inspired  by a
theoretical model, because it ``decouples''  the two power laws of the
model rotation  curve.  Our solution  to the central  density problem
likely  cannot solve  the  cuspy halo  issue  by itself  because
central  cusps are  more a  reflection of  the cold  and collisionless
properties of  the dark  matter than the  amount of  small-scale power
(\eg Ref. \cite{dekel}).  Nonetheless,  as we have already mentioned,
the cuspy halo issue is to some degree  degenerate with the central
density problem and it may be that solving the latter problem may go a
long way toward resolving the former.

A  third problem  associated with  the standard  LCDM paradigm  is the
dwarf satellite  problem \cite{DSP}.  In essence, this  problem can be
stated in the following way:  standard LCDM overpredicts the number of
satellite halos  with $10$ kms$^{-1} \lsim \Vmax  \lsim 50$ kms$^{-1}$
by as much as an order of magnitude relative to the number of observed
satellite  galaxies in  the  local group.   As  we mentioned  earlier,
Kamionkowski and Liddle  \cite{KL00} investigated solving this problem
with BSI initial power spectra.  It  is probable that at least part of
the  solution   lies  in  a  feedback   mechanism,  like  reionization
suppression \cite{BKW}.   However, the degree of  feedback needed will
depend crucially on the input  power spectrum.  We examine the subhalo
issue in  the context  of inflation in  a forthcoming  companion paper
\cite{ZB2}.   Briefly, we  find that  the discord  between  theory and
observation can  be greatly allayed  by considering models  similar to
those studied here and thus, the feedback needed to meet observations 
can be greatly reduced or even eliminated.

Related to the  dwarf satellite problem is the  recent result of Dalal
and Kochanek \cite{DK}.  The perturbing effect of
substructure in strong gravitational lenses allowed them to
constrain the fraction of the host halo mass
bound up in substructure  to be $0.006 \le f_{\rm
sat} \le 0.07$ (90\% confidence).
They used this result to limit  the tilt
of the primordial  spectrum and put constraints on the
neutrino mass, and they
obtained $n \ge 0.94$ and  $m_{\nu}  \le  0.74$ eV  at  95\%
confidence.  Our  results on substructure  differ from those  of Dalal
and  Kochanek.  As we discuss in our forthcoming paper
\cite{ZB2}, we  find that  for a  host halo  of 
the relevant mass, the total mass fraction in subhalos
is typically  larger than the  lower limit of  Dalal and
Kochanek  ($f_{\rm sat}  \ge  0.006$) even  with significantly  tilted
primordial  spectra, $n  \lsim 0.8$.   Thus the  tilt of  the
primordial power spectrum may  not yet be significantly constrained by
strong lensing results.  
However,  as we have demonstrated here,
the long ``lever arm'' from COBE scales to the subgalactic
regime offers a potentially useful avenue for constraining
models of inflation.  Perhaps future lensing studies 
will provide more significant limits, 
and thus  test the intriguing possibility that
galaxy rotation curves are telling
us something fundamental about the early universe.

\begin{acknowledgments}

This work benefitted from useful conversations with David 
Caldwell, Savvas Koushiappas, Joel Primack, Stuart Raby, Gary Steigman, Terry Walker and 
David Weinberg and was inspired by an email correspondence with Steen Hansen.  
We are grateful to Rob Swaters for making his rotation curves 
available to us and providing us with a copy of his thesis work.  
We thank Leszek Roszkowski for directing us toward useful references.  We 
thank Chris Power for pointing out an error in Eq. (18) in an earlier version 
of this work.  We would also like to thank an anonymous referee for making 
several suggestions that improved the quality of this paper.  We were supported 
by U.S. DOE Contract No. DE-FG02-91ER40690.
\end{acknowledgments}

\end{document}